\documentclass[12pt]{iopart}
\usepackage{graphicx}
\begin{document}
\newcommand{\cfourften}{C$_4$F$_{10}$}
\newcommand{\degC}{$^{\circ}$C}
\newcommand{\carbon}[1]{$^{#1}$C}
\newcommand{\flourine}{$^{19}$F}

\title{New Insights into Particle Detection with Superheated Liquids}
\author{The PICASSO Collaboration\\
S. Archambault$^a$\footnote[1]{present address: Department of Physics, McGill University, Montr\'{e}al, H3A 2T8, Canada} , F. Aubin$^{a}$\footnotemark[1], M. Auger$^a$\footnote[2]{present address: Laboratorium f\"ur Hochernergiephysik, Universit\"at Bern, CH-3012 Bern, Switzerland} , M. Beleshi$^b$, E. Behnke$^c$,  J. Behnke$^c$,  B. Beltran$^d$, K. Clark$^e$\footnote[3]{present address: Department of Physics, University of Oxford, OX1 3PU, Oxford, UK} , X. Dai$^e$\footnote[4]{present address: AECL Chalk River Laboratories, Chalk River ON, K0J 1J0, Canada} , A. Davour$^e$, F. Debris$^a$, J. Farine$^b$, M.-H. Genest$^a$\footnote[5]{present address: Fakult\"at f\"ur Physik, Ludwig-Maximilians-UniversitŠ\"at, D-85748 Garching, Germany} , G. Giroux$^{a}$\ddag, R. Gornea$^{a}$\ddag, R. Faust$^a$, H. Hinnefeld$^c$, A. Kamaha$^e$, C. Krauss$^d$, M. Lafreni\`{e}re$^a$, M. Laurin$^a$, I. Lawson$^f$ , C. Leroy$^a$, C. L\'{e}vy$^e$ \footnote{present address: Physik Department, Universit\"at M\"unster, D-48149, M\"unster, Germany}, L. Lessard$^a$, I. Levine$^c$, J.-P. Martin$^a$, S. Kumaratunga$^a$, R. MacDonald$^d$, P. Nadeau$^b$\footnote{present address: Department of Physics, Queens University, Kingston, K7L 3NG, Canada}, A. Noble$^e$, M.-C. Piro$^a$, S. Pospisil$^g$, N. Starinski$^a$, I. Stekl$^g$, N. Vander Werf$ ^c$, U. Wichoski$^b$, V. Zacek$^{a}$ \footnote {Corresponding author: \emph{E-mail address:} zacekv@lps.umontreal (V. Zacek)} }
\address{$^a$ D\'{e}partement de Physique, Universit\'{e} de Montr\'{e}al, Montr\'{e}al, H3C 3J7, Canada}
\address{$^b$ Department of Physics, Laurentian University, Sudbury, P3E 2C6, Canada}
\address{$^c$ Department of Physics \& Astronomy, Indiana University South Bend,
South Bend, IN 46634, USA}
\address{$^d$ Department of Physics, University of Alberta, Edmonton, T6G 2G7, Canada}
\address{$^e$ Department of Physics, Queens University, Kingston, K7L 3NG, Canada}
\address{$^f$ SNOLAB, 1039 Regional Road 24, Lively, P3Y 1N2, Canada}
\address{$^g$ Institute of Experimental and Applied Physics, Czech Technical University in Prague, Prague, Cz-12800, Czech Republic}

\begin{abstract}
We report new results obtained in calibrations of superheated liquid droplet detectors used in dark matter searches with different radiation sources (n,$\alpha$,$\gamma$). In particular, detectors were spiked with alpha-emitters located inside and outside the droplets. It is shown that the responses are different, depending on whether alpha particles or recoil nuclei create the signals. The energy thresholds for $\alpha$-emitters are compared with test beam measurements using mono-energetic neutrons, as well as with theoretical predictions. Finally a model is presented which describes how the observed intensities of particle induced acoustic signals can be related to the dynamics of bubble growth in superheated liquids. An improved understanding of the bubble dynamics is an important first step in obtaining better discrimination between particle types interacting in detectors of this kind.
\end{abstract}


\section{Introduction}
The PICASSO dark matter experiment uses the superheated droplet technique, which is based on the operation principle of the classic bubble chamber \cite{1,2,3,4,5}. Detectors of this kind are threshold devices, where the operating parameters (pressure and temperature) determine the energy threshold. Since each temperature, at a given pressure, corresponds to a defined recoil energy threshold, the spectrum of the particle induced energy depositions can be reconstructed in superheated liquids by measuring the rate as a function of temperature. 

In the case of PICASSO the active detector liquid is dispersed as droplets of a metastable superheated perfluorobutane, \cfourften, and the detectors are operated in a temperature range such that nuclear recoils in the keV range induced by interactions with Weakly Interacting Particles (WIMPS) could trigger bubble formation. These explosive evaporations are accompanied by acoustic signals, which are recorded by piezoelectric transducers.  
In previous studies the PICASSO collaboration showed that the acoustic signals contain information about the nature of the primary event \cite{6,7,8}: it was observed that the acoustic signals produced by alpha emitters are more intense than the signals of neutron or WIMP induced recoil events. Recently, this effect was confirmed by the COUPP and SIMPLE collaborations, which used it for a substantial background reduction in their dark matter searches \cite{9,10}.  

The underlying physics process can be explained by the hypothesis that in the case of alpha emitters the recoiling nucleus and the extended alpha track contribute with at least two vaporization centres to the total signal, whereas the signals of the much more localized nuclear recoils carry the imprint of one single nucleation only. In the following we describe recent efforts to consolidate this hypothesis. 

For this purpose PICASSO detectors were spiked with alpha emitters of known types and energies ($^{241}$Am and $^{226}$Ra) and the emitters were deliberately located outside and inside the droplets. The results, together with data obtained from existing detectors containing relatively large contaminations with alpha emitters, support the proposed model and show that mono-energetic nuclear recoils following alpha decays are detected at lower temperatures than alpha particles. The energy depositions of nuclear recoils are larger than the energy deposition by the Bragg peak of the alpha particles themselves. This latter information can then be used to infer more precise estimates for the parameters of the model proposed by Seitz \cite{11}, which still serves as the reference theory to describe the radiation sensitivity of superheated liquids. The results from measurements using alpha emitters are then compared with results from test beam measurements using mono-energetic neutrons with energies between 4.8 keV and 4 MeV and are found to be in good agreement. Data for gamma ray induced nucleations also fit the described model well. Finally a plausible argument is presented to explain the observed alpha-recoil discrimination in terms of the dynamics of bubble growth in superheated liquids. This is largely uncharted terrain, nevertheless some conclusions can be drawn which shed light on the early phase of particle induced bubble formation.

\section{Detection Principle and Theoretical Model}\label{sec:model}
 For a phase transition to occur in a superheated liquid the prevailing theoretical model proposed by Seitz \cite{11} predicts that a critical minimum amount of energy $E_c$ has to be supplied within a local thermal spike and if the resulting proto-bubble reaches a volume of critical radius $R_c$, it becomes thermodynamically unstable and grows rapidly. Thermodynamics predicts that the growth of the bubble passes through several stages of acceleration and deceleration, which also gives rise to a detectable pressure wave (sect.10). 

Both, $R_c$ as well as $E_c$ decrease exponentially with temperature and are given by

\begin{equation}
	R_c(T)=\frac{2\sigma}{\Delta p}\label{eq:1}
\end{equation}
\begin{equation}
E_c(T) = - \frac{4\pi}{3} R_c^3 \Delta p + \frac{4\pi}{3} R_c^3\rho_v h_{\mathrm{lv}} + 4\pi R_c^2\left(\sigma-T\frac{d\sigma}{dT}\right)+W_{\mathrm{irr}}  \label{eq:2}
\end{equation}

\noindent where $\sigma$  is the surface tension at the liquid-vapor interface,   $\Delta p = p_v - p_e$ is the degree of superheat, which is the difference between the vapour pressure $p_v$ and the external pressure $p_e$, $\rho_v$ is the density of the gas phase, and $h_{\mathrm{lv}}$ is the latent heat of evaporation.  All of these quantities depend on the temperature of operation T. The first term in expression (\ref{eq:2}) is the reversible work done during expansion to a bubble of size $R_c$ against the pressure of the liquid. The second term represents the energy needed to evaporate the liquid and is necessary to be provided if the subcritical bubble grows faster than energy can be supplied from its liquid vicinity. The third term describes the work needed to create the liquid vapour interface of the proto-bubble. $W_{\mathrm{irr}}$ is the work which goes into irreversible processes, like acoustic wave emission; this contribution is small compared to the others.

Radiation induced phase transitions imply that the locally deposited kinetic energy of a traversing particle exceeds the critical energy, i.e. $E_{\mathrm{dep}}  \ge E_c$ (T) and that the stopping power of the particle is large enough to supply this energy as heat over a distance $L_c$ such that it is effective in reaching the critical energy within a protobubble of critical size $R_c$ \cite{12}:

\begin{equation}
E_{\mathrm{dep}} (T) = \int_0^{L_c(T)}\frac{dE}{dx}dx \ge E_c(T)
\end{equation}\label{eq:3}

 Experimentally relatively good agreement is reported between $E_{\mathrm{dep}}$ and $E_c$ for several halocarbons and for energies around $E_c$ = 17 keV and larger \cite{13}. The situation is different for $L_c$, where a large range of values is quoted. Since $R_c$ is the natural length scale of the process, $L_c$ is usually given in terms of $L_c = b R_c$, where $b$ varies from author to author:  $L_c$ = 2 $R_c$ appears to be intuitively justified and is supported by data for low energy thresholds below 20 keV;  $L_c = 2\pi R_c = 6.28 R_c$ is proposed in \cite{14} following arguments of stability of vapour jets in liquids, whereas some authors propose $L_c = (\rho_v/\rho_l)^{1/3}R_c \approx 6.6 R_c$, but also values up to 18 have been reported \cite{12,13,14,15,16,17,18,19}.  As described in section \ref{sec:detLc}, our recent results from calibration measurements with mono-energetic neutrons and alpha emitters in PICASSO shed additional light on this controversial issue. 

The active detector material in PICASSO, \cfourften, has a boiling temperature of $T_b = -1.7 ^{\circ}$C at a pressure of 1.013 bar and a critical temperature of $T_c = 113.3 ^{\circ}$C; therefore at ambient temperature and pressure this liquid is in a metastable, superheated state. Table \ref{tab:1} summarizes the predictions of the Seitz model for $E_c(T)$ and $R_c(T)$ using relations (\ref{eq:1}) and (\ref{eq:2}) and values for $\rho_v$, $h_{\mathrm{lv}}$, $\sigma$ and $p_v$ compiled by NIST \cite{20}. The contribution of $W_{\mathrm{irr}}$ in relation (\ref{eq:2}) is small and has been neglected. The ranges of the recoiling \carbon{12} and \flourine{} atoms, $R_C$ and $R_F$, are also included in Table \ref{tab:1} for comparison at the quoted values for the threshold energies $E_c(T)$. 
\begin{table}
\begin{center}
\begin{tabular}{|ccccc|}
\hline
T (\degC{})&	$E_c$ (keV)&	$R_c$ (nm)&	$R_F$(nm)&	$R_C$(nm)\\
\hline
10&	1000	&280&	2540&	3600\\
20&	111.1&	111&	590&	1090\\
30&	27.8&	60&	99&	210\\
40&	7.45&	35.3&	20&	40\\
\hline
\end{tabular}
\end{center}

\caption{Predictions by the Seitz model for $E_c$(T) and $R_c$(T) in \cfourften using relations (\ref{eq:1}) and (\ref{eq:2}) and values for $\rho_v$, $h_{\mathrm{fg}}$, $\sigma$ and  $p_v$ compiled by NIST \cite{20}. Also quoted are the ranges $R_{\mathrm{F,C}}$ for fluorine and carbon ions with kinetic energies corresponding to the given values of $E_c$.}\label{tab:1}

\end{table}

It is interesting to note that at e.g. $20 ^{\circ}$C most of the work required to create a critical bubble is spent in evaporating the liquid, $W_{\mathrm{ev}} = 80$ keV, the work required to create the liquid vapour interface amounts to $W_{\mathrm{lv}}$ = 36 keV and the mechanical work done during bubble expansion amounts to $W_m = -$ 4.6 keV. These contributions are affected by errors at the level of 15 -- 20\% due to uncertainties in the thermodynamic parameters (and these uncertainties increase with temperature).

\begin{figure}
\begin{center}
 \includegraphics[scale=0.6]{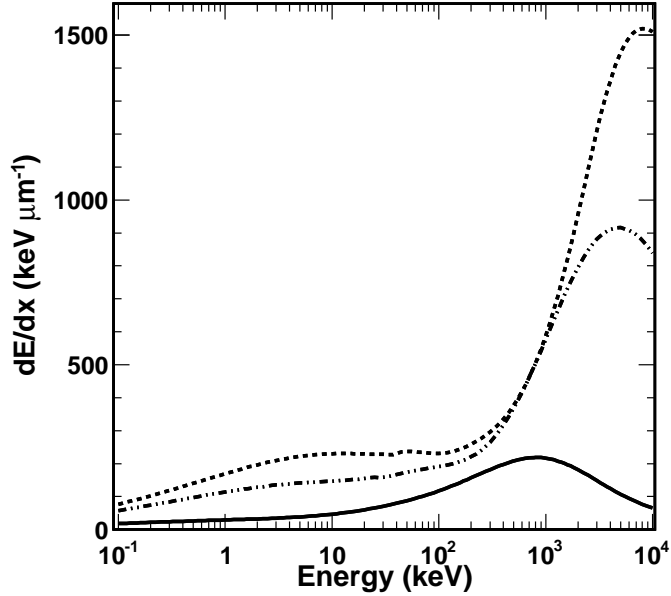}
\end{center}
\caption{Stopping power in keV/$\mu$m for alpha particles (continuous), fluorine nuclei (dotted) and carbon nuclei (dash-dotted) in \cfourften \ calculated with SRIM \cite{21}. In the energy range below 500 keV fluorine always has the higher dE/dx.}\label{fig:1}
\end{figure}

The specific energy losses in liquid \cfourften \ of the particles used in the calibration measurements discussed below, i.e alpha particles and neutron induced fluorine and carbon recoil nuclei were calculated with SRIM \cite{21}; the results are shown in Figure \ref{fig:1}. For the energy range considered here, with E$_{F,C,\alpha} <$  800 keV, fluorine always has the higher stopping power, followed by carbon and alpha particles. The stopping power at the Bragg peak of alpha particles almost equals the stopping power of fluorine below 100 keV. 

\section{Detectors and Read Out}\label{sec:det}
The current PICASSO detector generation consists of cylindrical modules of 14 cm diameter and 40 cm of height \cite{5}. They are fabricated from acrylic and are closed on top by stainless steel lids sealed with polyurethane O-rings. Each detector is filled with 4.5 litres of polymerized emulsion loaded with droplets of \cfourften; the droplet volume distribution peaks at diameters of around 200$\mu$m. The active mass of each detector is typically around 85 g. The active part of each detector is topped by mineral oil, which is connected to a hydraulic manifold. After a measuring cycle the detectors are compressed at a pressure of 6 bar in order to reduce bubbles to droplets and to prevent bubble growth which could damage the polymer. The operating temperature of the modules is controlled with a precision of $\pm$0.1$^{\circ}$C.      
Each detector is read out by nine piezo-electric transducers. Three transducers are distributed around the detector at each of three different heights. They are flush mounted on a flat spot milled into the acrylic. The transducers are ceramic disks (Pz27 Ferroperm) with a diameter of 16 mm and 8.7 mm thickness and a sensitivity of 27 $\mu$V/$\mu$bar.  The piezoelectric sensors are read out by custom made low-noise preamplifiers that serve a double purpose: providing impedance conversion and strong amplification (gain x 3000 between 0.5 kHz and 130 kHz). The bandwidth of the amplified signal is limited to the range of 1 to 80 kHz using a series of customizable RC filters. The amplified output signal is digitized using a 12-bit analog to digital converter (ADC) with serial output. The maximum amplitude of the digital signal is  2 V. 

The trigger threshold is individually set for each channel, normally at $\pm$300 mV. The  trigger condition requires at least one signal from a detector module crossing the threshold, and in this case will initialize the readout of all channels from this module. The system stores 8192 samples at a sampling frequency of 400 kHz, with 1024 samples before the trigger time. The total recorded signal is 20.48 ms long.                        

\section{Energy Calibration with the Alpha Emitters $^{241}$Am and $^{226}$Ra} \label{sec:calalpha}
\begin{figure}
\begin{center}
 \includegraphics[scale=0.7]{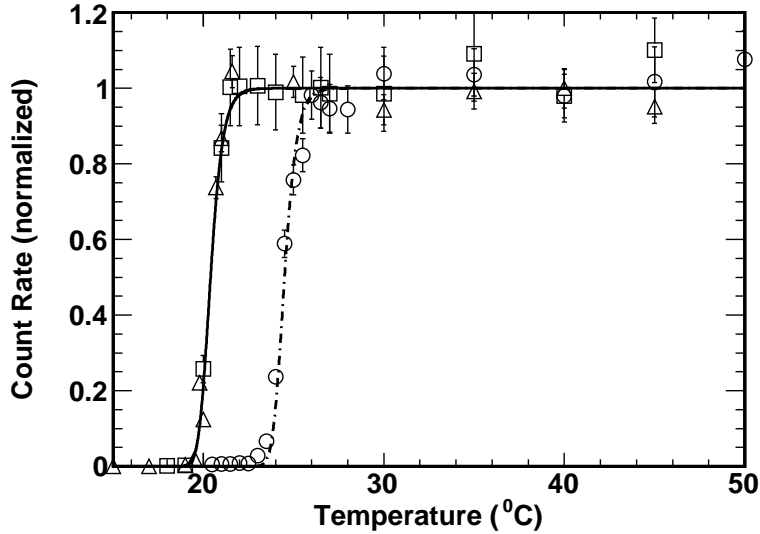}
\end{center}
\caption{Response of detectors spiked with the alpha-emitters $^{241}$Am and $^{226}$Ra. The curve with the higher threshold temperature was obtained after spiking the detector matrix with $^{241}$Am (circles) and only $\alpha$-particles entering the droplets can induce nucleation. At threshold $\alpha$-particles trigger nucleation with energy depositions at the Bragg peak. The lower threshold was obtained with two $^{226}$Ra spiked detectors (squares and triangles). Here the recoiling $^{210}$Pb nucleus with the highest recoil energy in the decay chain ($E_{rec}$ = 146 keV) defines the thereshold. The respective count rates of the different measurements are normalized at the plateau.}\label{fig:2}
\end{figure}

Two detectors were prepared especially to study the response of superheated liquids to alpha decays of known origin. In one detector, the polymer in which the droplets are suspended was first spiked with an aqueous solution of $^{241}$AmCl with an activity of 25.56 Bq and after completion of the measurements this same detector was spiked with 10 Bq of $^{226}$Ra. 
The other detector was exclusively spiked with 10 Bq of $^{226}$Ra. Both  detectors were shortened versions,  1/3 in height of normal PICASSO modules. They were read out in one horizontal transducer plane, i.e. by three  piezoelectric sensors arranged under 120$^{\circ}$ with respect to each other and mounted on the outside of the detector wall.  The observed count rates as a function of temperature are shown in Figure \ref{fig:2}. 

$^{241}$Am decays follow the reaction $^{241}$Am$\, \to\, ^{237}$Np + $\alpha$ + 5.64 MeV. The count rate of the $^{241}$Am-spiked detector exhibits a threshold at 22$^{\circ}$C, reaches a flat plateau at 26$^{\circ}$C and  traverses 50\% of the plateau rate at 24.5$^{\circ}$C. This temperature corresponds to an energy threshold of $E_{\mathrm{th}}$ (24.5$^{\circ}$) = 71 keV according to neutron calibrations (sect. \ref{sec:calibmono}). Since the alpha emitters are predominantly located outside the droplets (sect. 9.2), the only alpha particles which can trigger a phase transition at threshold are those with the highest specific energy deposition, which corresponds to the Bragg peak.  At higher temperatures the liquid becomes sensitive to smaller dE/dx on the tracks, but since the detector is already fully sensitive to alpha particles immediately above threshold, the temperature response levels off to a plateau. 

After completion of data taking with the $^{241}$Am spike, the same detector was loaded with $^{226}$Ra, this time by injecting locally with a syringe an aqueous solution of RaCl (10 Bq).  $^{226}$Ra (T$_{1/2}$ = 1602 y) decays into $^{222}$Rn (T$_{1/2}$ = 3.8d), which then decays following the sequence of transitions  $^{222}$Rn$ \,\to\, ^{218}$Po$ \,\to\, ^{214}$Pb$\,\to\, ^{214}$Bi$\,\to\, ^{214}$Po$\,\to\, ^{210}$Pb via three $\alpha$- and two $\beta$- decays; the rest of the chain is too long-lived to be relevant here.  The energies of the emitted $\alpha$-particles are 5.49 MeV, 6.0 MeV and 7.69 MeV, respectively, and the half lives are short with respect to the $^{226}$Ra half life (3.8 d, 3 min, 27 min, 19 min, 0.2 ms).  By rate measurements and visual inspection of the bubbles formed, the radon induced $\alpha$-activity could be observed to diffuse slowly within a couple of days over the entire detector volume, including the \cfourften \ droplets themselves. When equilibrium for this portion of the chain was reached, data were taken as a function of temperature and the results are shown in Figure \ref{fig:2}.      
        
The observed threshold is now shifted by about 4$^{\circ}$C towards lower temperatures, i.e larger energy depositions: it starts at 19$^{\circ}$C and reaches its plateau at 22$^{\circ}$C. However, the observed shift cannot be attributed to the energy deposition of the $\alpha$-particles emitted by the Ra-chain, since they have the same maximum stopping power at the Bragg peak than those emitted in $^{241}$Am-decay; rather the reduced energy threshold is now caused by $^{210}$Pb nuclei recoiling inside the droplets with energy of 146 keV, since this is the nucleus with the highest recoil energy in the chain. 

By further raising the temperature, the detector becomes subsequently sensitive to the lower energetic $^{214}$Pb (112 keV) and $^{218}$Po (101 keV) recoil nuclei. First $^{214}$Pb recoils add to the observed count rate, but once the energy threshold is low enough to allow $^{218}$Po recoils to trigger, the $^{214}$Pb recoils, which follow $^{218}$Po decays (T$_{1/2} \approx$ 3 min.), are no more able to contribute several minutes after run start. This is due to the fact that once a phase transition has occurred in a droplet it is no longer sensitive to subsequent energy depositions.  The same situation arises for $^{214}$Po decays which are gradually rendered undetectable following $^{222}$Rn/$^{218}$Po decays in the same droplet for measuring times exceeding the half lives of the two beta decays after run start (T$_{1/2}$ = 27 min and 19 min).  This expected asymptotic decrease in count rate due to the depletion of $^{214}$Po decays has been observed for temperatures T $>$ 20$^{\circ}$C and for measuring times lasting up to two hours after run start. In order to eliminate this time dependence, the count rates shown in Figure \ref{fig:2} were always calculated for the same time interval after run start. 

The observed threshold curve is characterized by a steep slope, but the data are not precise enough to reveal a step-like substructure which should be caused by the 34 keV energy difference between the $^{214}$Pb and $^{210}$Pb recoils. However, the 146 keV recoil energy of the $^{210}$Pb nuclei at the observed threshold temperature is consistent with the threshold obtained for neutron induced \flourine{} recoils discussed in sect. \ref{sec:calibmono}. By increasing the temperature from 22$^{\circ}$C to 25$^{\circ}$C the $^{226}$Ra spiked detector becomes also sensitive to $\alpha$-particles, but since the detector is already fully sensitive, the count rate remains unaffected.  However, above 25$^{\circ}$C, $\alpha$-particles contribute to the amplitude of the acoustic signal. This effect will be described in sect. \ref{sec:acoustic}.  

Although the $^{241}$Am activity is still present after the $^{226}$Ra spike, it does not show up as a step at 22$^{\circ}$C in Figure 2. This is due to the fact that after the  Ra-spike the recorded count rate increased by a factor of 110 \footnote[1]{This does not appear in Figure 2 since all rates are normalized to a common plateau value}. This strong increase in count rate is not yet completely understood. It could possibly be attributed to the combination of two effects: 1) the larger geometric detection efficiency for alpha decays within the droplets and 2) an additional concentration of alpha emitters in the droplets caused by an increased solubility of $^{222}$Rn in fluorocarbons like \cfourften \ compared to the $^{222}$Rn solubility in the water based polymer.

The second detector used in this study, similar in size and composition, but spiked exclusively with 10 Bq $^{226}$Ra, reproduced the above described threshold results (Figure \ref{fig:2}). Also its count rate, normalized to the active mass of \cfourften \, was found comparable to that of the Am loaded  detector after the Ra spike \cite{22}. Another, similar study of alpha emitters  was described by Hahn \cite{23} employing $^{238}$U and $^{232}$Th spikes in CCl$_2$F-CClF$_2$ operated under negative pressure. In those measurements, recoils from $^{210}$Po $\alpha$-decays could be clearly separated from the 67 keV smaller recoil energies produced in $^{212}$Po decays.

\section{Energy Calibration with Mono-Energetic Neutrons}\label{sec:calibmono}
The dependence of the threshold energy $E_{\mathrm{th}}$ on temperature and pressure was studied with neutron induced nuclear recoils. For this purpose extensive calibrations were performed at the Montreal Tandem van de Graaf facility with mono-energetic neutrons ranging from $E_n$ = 4.8 keV up to 4 MeV. In the case of nuclear recoils induced by neutrons of low energy the interaction is mainly through elastic scattering on fluorine and carbon nuclei. Inelastic collisions only occur if the centre-of-mass kinetic energies of the neutrons are  higher than the first excitation level of the nuclei (1.5 and 4.3 MeV for \flourine{} and \carbon{12}, respectively). Absorption of neutrons by the   \flourine{} nucleus followed by alpha particle emission requires a neutron energy of 2.05 MeV. 

Assuming neutron elastic scattering on nuclei, the recoil energy, E$^i_R$, of the nucleus i is given by 
 
\begin{equation}
E^i_{R} = \frac{2m_n m_{N_i} E_n(1-\cos\theta)}{(m_n + m_{N_i})^2}
\end{equation}

\noindent where $E_n$ and $\theta$ are the incident neutron energy and the neutron scattering angle in the centre-of-mass system, $m_n$ and $m_{N_i}$ are the masses of the neutron and the nucleus $N_i$, respectively. At a given neutron energy $E_n$ the recoiling nuclei are emitted with an angular distribution, every angle being associated to a specific recoil energy ranging from 0 keV at $\theta$ = 0 up to the maximum energy $E^i_{R,max}$  for  $\theta$ =180$^{\circ}$.  At the small energies considered here, the angular distribution is isotropic in the centre-of mass system and the recoil energy spectrum dR$^i$/dE$^i_R$ has a rectangular, boxlike shape up to E$^i_{R,max}$:

\begin{equation}
	E^i_{R, max}   = f_i E_n = \frac{4m_n m_{N_i}E_n}{(m_n+m_{N_i})^2} 
\end{equation}

\noindent The factor $f_i$ gives the maximum fraction of the energy of the incident neutron transmitted to the nucleus i, where $f_F$ = 0.19 and 0.28 for \flourine{} and \carbon{12}, respectively. 

The mono-energetic neutrons used for calibration were produced via nuclear reactions with mono-energetic protons via the $^7$Li(p,n)$^7$Be and $^{51}$V(p,n)$^{51}$Cr reactions. The measurements with the Li target (10 $\mu$g/cm$^2$) cover a range of neutron energies from 100 keV up to 4 MeV and the results obtained are discussed in detail in \cite{18}.  With improved proton beam stability these calibrations were recently extended in PICASSO with a $^{51}$V target (9 $\mu$g/cm$^2$) down to 4.8 keV.  In order to acquire sufficient statistics close to threshold, the proton beam energies were tuned to individual resonances in the $^{51}$V(p,n)$^{51}$Cr reaction cross section \cite{24}.  In particular the five resonances quoted in Table \ref{tab:2} have been selected, each of which has an intrinsic width below one keV.
\begin{table}
\begin{center}
\begin{tabular}{|c|c|c|}
\hline
Resonance&	$E_p$ (MeV)&	$E_n$ (keV)\\
\hline
I	&1.568&	4.8\\
V	&1.598&	40\\
VII	&1.607&	50\\
VIII	&1.617&	61\\
XI	&1.651&	97\\
\hline
\end{tabular}
\end{center}
\caption{Five of the resonances of the $^{51}$V(p,n)$^{51}$Cr reaction used for neutron calibration. All five resonances have sub-keV intrinsic widths \cite{24}.}\label{tab:2}
\end{table}

The detectors used for these measurements are smaller in size (63 mL) with an active mass of 1g of \cfourften, but fabricated in the same way and with similar droplet dimensions as the standard 4.5L detectors. For each of the selected neutron energies, data were taken at $\theta \approx$   0$^{\circ}$ with respect to the beam, while ramping the temperature up and down for a given pressure. Since close to threshold the cosmic ray induced n-background can amount to 50\% of the total count rate, each neutron run at a fixed temperature was  followed by a background run at the same temperature (after 8h of recompression). 

For a fixed neutron energy the data have been normalized by the integrated proton current and the count rate of a $^3$He counter mounted behind the target was used to compensate for short off-resonance beam energy fluctuations \cite{25}. The measurements at the lowest neutron energy (4.8 keV) were particularly challenging since at threshold and above, the detectors had to be operated between 48 and 60\degC{} where \cfourften \ becomes sensitive to the 320 keV gamma rays (T$_{1/2}$ = 28 d) following de-excitation of $^{51}$Cr. Therefore this background had to be measured independently during a beam off period after each neutron run and subtracted from the data.   

\begin{figure}
\begin{center}
 \includegraphics[scale=0.3]{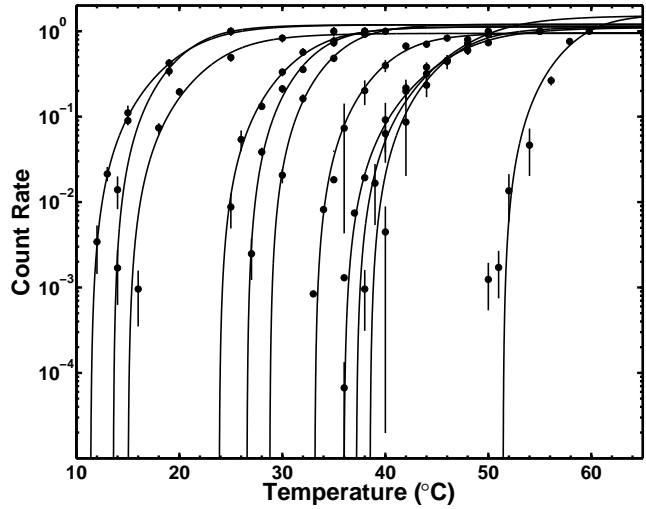}
\end{center}
\caption{Detector response to mono-energetic neutrons as a function of temperature (\degC{}). The detectors were 10 and 15 mL modules loaded with \cfourften \ doplets. From left to right the threshold curves correspond to neutron energies of 4 MeV, 3 MeV, 2 MeV, 400 keV, 300 keV, 200 keV, 97 keV, 61, keV, 50 keV, 40 keV and 4.8 keV, respectively. The five lowest energies were obtained from resonances of the $^{51}$V(p,n)$^{51}$Cr reaction,  the higher energy neutrons were produced with the $^7$Li(p,n)$^7$Be reaction; several  more energies obtained with the Li-target are shown in Figure \ref{fig:4}. The curves shown are fits which include attenuation and resolution effects from simulations.}\label{fig:3}
\end{figure}

\begin{figure}
\begin{center}
 \includegraphics[scale=0.6]{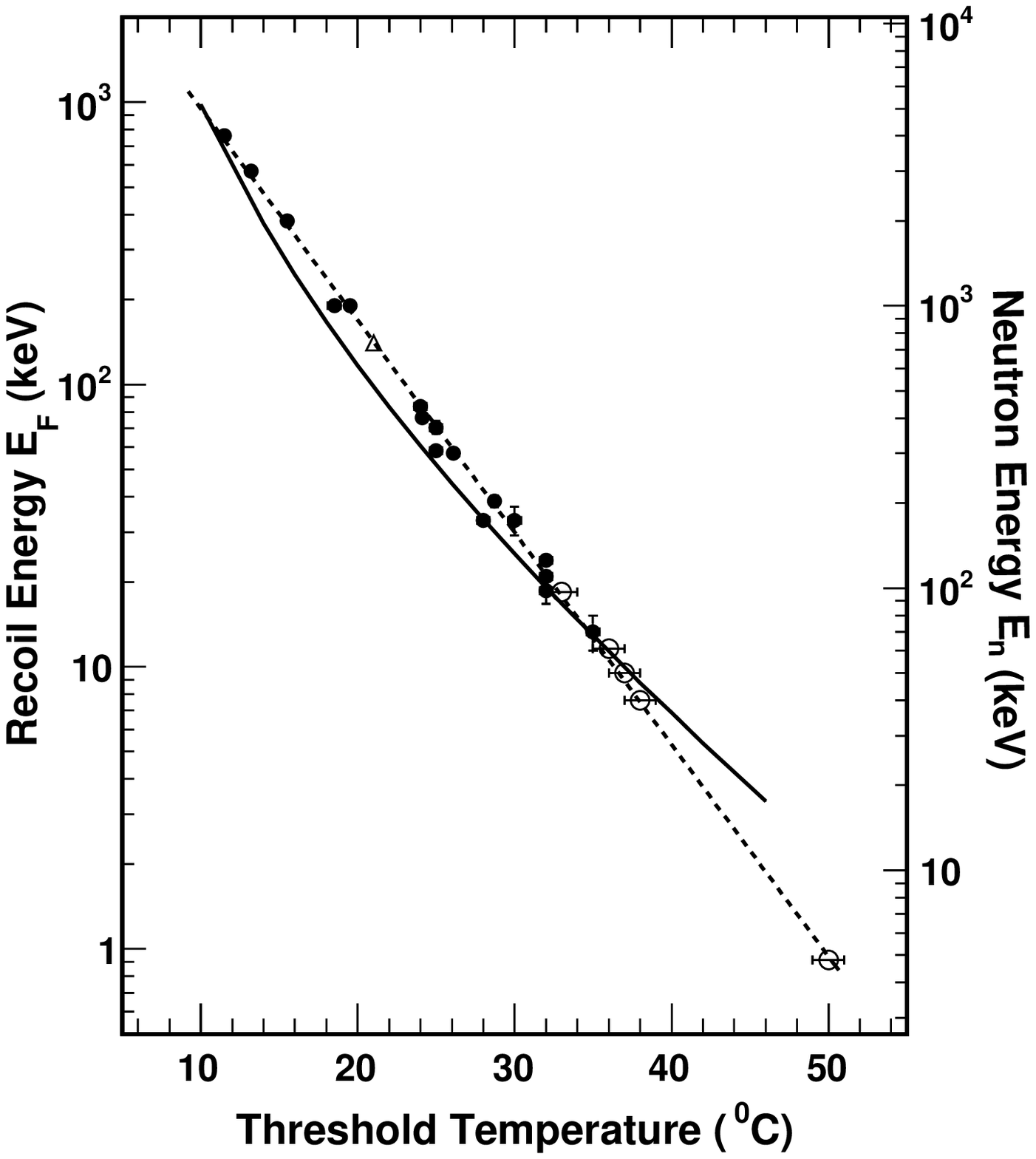}
\end{center}
\caption{Relation between the energies of mono-energetic neutrons (right vertical scale) and the temperature measured at threshold (Figure 3); the left vertical scale displays the maximum recoil energy of fluorine, which corresponds to the selected neutron energy. For fluorine recoil energies from E$_F$ = 0.9 keV to 760 keV, the data are well described by an exponential dependence on temperature (broken line). The open triangle at 21\degC{} corresponds to the energy deposition of 146 keV of $^{210}$Pb recoils following $^{222}$Rn decay (Figure \ref{fig:2}). The continuous line represents E$_c$(T) calculated in relation (\ref{eq:2}); it  is  the work required to create a bubble with critical radius R$_c$(T). 
}\label{fig:4}
\end{figure}

The recorded count rates for the Li and V targets at different neutron energies at ambient pressure and as a function of temperature are compiled in Figure \ref{fig:3}. From these measurements, the threshold temperature, $T_{\mathrm{th}}$, can be extracted for a given neutron energy by fitting the data to a theoretical response function including energy losses in the target and the detector matrix and by keeping the intrinsic energy resolution as a free parameter (sect. \ref{sec:eres}).  From this the neutron threshold energy as a function of temperature can be inferred as is shown in Figure \ref{fig:4}. For the energies considered here, $E^n_{th}$(T) follows an exponential dependence on temperature. A similar exponential trend was observed by other authors for a series of halocarbons such as C$_4$F$_8$, CCl$_2$F$_2$, C$_2$H$_3$ClF$_2$ and C$_2$Cl$_2$F$_4$ \cite{13}. 

In practice it is more interesting to know the threshold of the minimum   nuclear recoil energy at a given temperature.  Due to the composition of the target in use, \cfourften, there are two possibilities in converting neutron energy into recoil energy: 1) assuming that the response depends on the energy which is deposited on the entire recoil track, then the threshold should be attributed to the more energetic carbon nucleus; 2) if the recoil nucleus with the greater dE/dx triggers, then it is fluorine that defines the threshold. In both cases the energy thresholds (at 1 bar) are obtained from the fit to the neutron threshold data and in the case of \flourine{} follow the relation: 

\begin{equation}
E^F_{th}(T) = 0.19 E^n_{th} = (4.93 \pm 0.15) \times 10^3 \exp(-0.173 \;T(^{\circ}C)) (keV)
\end{equation}
\label{eq:6}

\noindent The kinematic factor 0.19 relates the measured neutron threshold energy to the maximum respective nuclear recoil energy; if \carbon{12} would trigger the kinematic factor would be 0.28. The error of 3\% is largely due to the systematic errors of  $\pm$ 0.2\degC{} in the temperature measurement during the test beam runs. Given the temperature range of operation in PICASSO, this translates in the case of \flourine{} recoils into a range of sensitivity from E$_F$  $>$ 2.0 keV at 45\degC{}   to E$_F$ $>$ 200 keV at 18.5\degC{}, respectively.  

Several observations support the conclusion that indeed fluorine atoms with their higher dE/dx are responsible for defining the threshold: 1) the neutron scattering cross section on \flourine{} is four to six times larger than the cross section on \carbon{12} in the considered energy range with substantial enhancement in its resonances. If carbon would trigger first, a second threshold should be observed in the threshold curves of Figure \ref{fig:3} at higher temperatures when scattering on $^{19}$F sets in; 2) the detection threshold for the 146 keV recoil$^{210}$Pb nuclei following alpha decays of $^{214}$Po matches the one for \flourine{} at the same energy; 3) for recoil energies between 10 keV and 1 MeV, the critical energy $E_c$(T) for bubble formation predicted by the Seitz model follows closely the $^{19}$F threshold; 4) measurements with C$_4$F$_8$ and C$_2$Cl$_2$F$_4$ show that the recoil energy of the species with the higher dE/dx matches well the predicted $E_c$(T) \cite{13,16}. 

             For threshold energies smaller than 10 keV, data and theory deviate from each other. The cause of this discrepancy is not yet understood, especially since it was observed for halocarbons that $E_c$ approaches zero if the temperature attains 90\% of the critical energy $T_c$, which would correspond to 74$^{\circ}$C in the case of \cfourften \cite{18}; this temperature is also very close to the limit of stability observed during measurements of the $\gamma$-sensitivity discussed in sect. 6.  The fact that $E_c$(T) curves up at high temperatures might be due to a still incomplete description of the underlying processes once the critical radius $R_c$ approaches the nanometer scale. Thermodynamics also requires that the threshold curve bends up towards infinity at the boiling temperature $T_b$ = -1.7 $^{\circ}$C.

\section{Energy Response to Gamma Rays}\label{sec:gamma}
In contrast to the energy depositions of recoiling nuclei and alpha particles the main interaction process of $\gamma$-rays with the detector material occurs via Compton scattering. Because of their very small stopping power, recoiling Compton electrons cannot trigger a phase transition directly in the normal temperature range of operation. Rather the observed sensitivity to gamma rays is attributed to $\delta$-rays or Auger electrons scattered randomly along the tracks of the Compton electrons. These low energy electrons curl up at the end of their trajectory into highly localised clusters of ionisation or hot spots which rarely lead to energy depositions at the keV level. In particular it was found in simulations that the $\delta$ -ray energy spectra on tracks of electrons from 5 keV up to 500 keV, and on the tracks of 1 GeV muons are very similar in shape and  50\% of the emitted $\delta$ rays were found to deposit energies smaller than 0.05 keV \cite{18, 26}. Calibrations with $\gamma$-rays can therefore give only information about the probability distribution of clusters of energy on the tracks of Compton electrons.  

\begin{figure}
 \begin{center}
\includegraphics[scale=0.7]{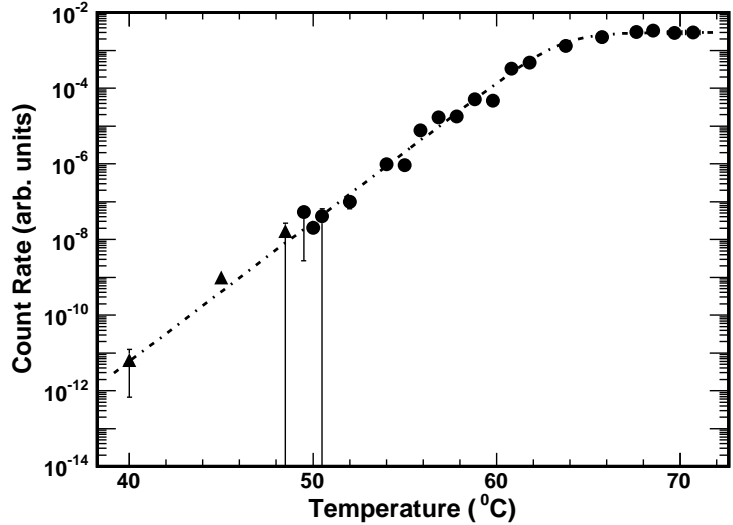}
\end{center}
\caption{Gamma-ray sensitivity as a function of temperature (\degC{}). The data at the four lower temperature points were recorded with a standard 4.5L detector (triangles); the higher temperature data were taken with a small 10 mL detector.  At the plateau the gamma detection efficiency is determined by the geometry and the probability that a Compton electron hits a droplet. Over 10 orders of magnitude in count rate the data are well reproduced by the sigmoid function described by relation (\ref{eq:7}).  After 72$^{\circ}$ the detector becomes sensitive to spontaneous nucleation.
}\label{fig:5}
\end{figure}

Such a study was performed with a $^{22}$Na source (0.7 $\mu$Ci), which yields 1.275 MeV $\gamma$-rays, as well as the two 511 keV photons from e$^+$e$^-$ annihilation. Compton scattering of the 1.275 MeV $\gamma$-rays produces recoiling electrons in the detector medium with an average energy of 500 keV and with a range of 1.1 mm, whereas the annihilation photons create electrons of 170 keV. Two different detectors were used: a standard 4.5L detector with an active mass of 78.9 $\pm$ 8 g was used to explore the low temperature response from 40 to 50$^{\circ}$C and a 10 mL detector with an active mass of 30 mg was used to measure temperatures from 48 up to 72$^{\circ}$C \cite{26, 27}.  The count rates as a function of temperature are shown in Figure \ref{fig:5} and it was found that the measured sensitivity can be fit over more than 10 orders of magnitude with a sigmoid function

\begin{equation}
\epsilon_{\gamma} = \frac{\epsilon_0}{1+\exp(\frac{T_0-T}{\tau})}
\end{equation}\label{eq:7}

\noindent with $T_0$ = 63.6\degC{} and $\tau$ = 1.78\degC{}. At the plateau, the measured efficiency ($\sim$5\%) corresponds roughly to the geometric efficiency that a Compton electron hits a droplet ($\sim$1\%).       
Measurements with other sources ($^{57}$Co, $^{60}$Co, $^{137}$Cs) showed that the response curves are identical and that the plateau efficiencies are comparable for $\gamma$-energies from 127 keV to 1.3 MeV.  This is expected, since for an average Z of the detector material of Z $\approx$ 5.5,  the Compton scattering cross section dominates in the energy range from 400 keV to 5 MeV and the mass attenuation coefficient does not vary much. In addition, the stopping power of the scattered electrons does not vary much in this region and since the $\delta$-ray production probability is proportional to the dE/dx on the particle track,  the  $\gamma$-detection efficiency reflects the constant  production efficiency of $\delta$-electrons. This scenario was confirmed by simulations, which in addition show that the sigmoid shape of the observed response curve can be reproduced well \cite{18, 26}.

Above 72\degC{} the detector becomes sensitive to spontaneous nucleation. An increase in temperature by 1$^{\circ}$ leads to an increase in count rate by approximately three orders of magnitude.
          
\section{Energy Resolution} \label{sec:eres}
By inspection of the alpha data in Figure 2 it is apparent, that the detection threshold is a well defined, but not a sharp step function; the count rate increases steeply, but gradually from threshold to full efficiency. The probability $P(E_{\mathrm{dep}}$, $E_{\mathrm{th}}$), that an energy deposition $E_{\mathrm{dep}}$ larger than the energy threshold $E_{\mathrm{th}}$ will generate a nucleation can be approximated by  

\begin{equation}
R(E_{\mathrm{dep}},E_{th}(T)) = 1 - \exp[a(1-\frac{E_{\mathrm{dep}}}{E_{\mathrm{th}}(T)})]
\end{equation}\label{eq:8}

\noindent where the parameter $a$ describes the observed steepness of the threshold: the larger $a$  is, the sharper the threshold is defined. This parameter is related to the intrinsic energy resolution and reflects the statistical nature of the energy deposition and its conversion into heat \cite{28}. It has to be determined experimentally for each superheated liquid and for different particle species, respectively.  Our measurements with alpha emitters with well defined, mono-energetic recoil nuclei ($^{210}$Pb)  indicate a threshold which can be described best with $a$ = 10 $\pm$ 1 at 146 keV; alpha particles depositing their energy at the Bragg peak seem to produce a somewhat less steep threshold described by  $a$ = 5.8 $\pm$ 0.7 at 65 keV (Figure 2).  

This parameter is more difficult to extract from calibrations with mono-energetic and poly-energetic neutrons due to the continuous spectral distributions of the recoiling nuclei. Our data with mono energetic neutrons above 400 keV are compatible with $a$  = 10 $\pm$ 5, at lower energies smaller values appear favoured with $a$  = 2.5 $\pm$ 0.5 \cite{29}.  A more precise study of a suspected temperature dependence of the resolution parameter is the subject of ongoing measurements.

\section{Determination of the Critical Length $L_c$}\label{sec:detLc}
 	           The relatively good agreement between deposited energy at threshold $E_{\mathrm{dep}}$ and the critical energy  $E_c$ required for nucleation below 40$^{\circ}$C observed in neutron calibrations allows an estimate of the effective ion track length $L_c = b\, R_c$ over which the energy deposition is distributed. We follow here the model proposed by d'ÕErrico, which assumes that the vapour cavity formed initially may extend along the charged particle track, before ending up in at least one structure of spherical shape of size $R_c$ \cite{2}. 
									    
\subsection{L$_c$ from alpha emitters}  
\begin{figure}
\begin{center}
 \includegraphics[scale=0.6]{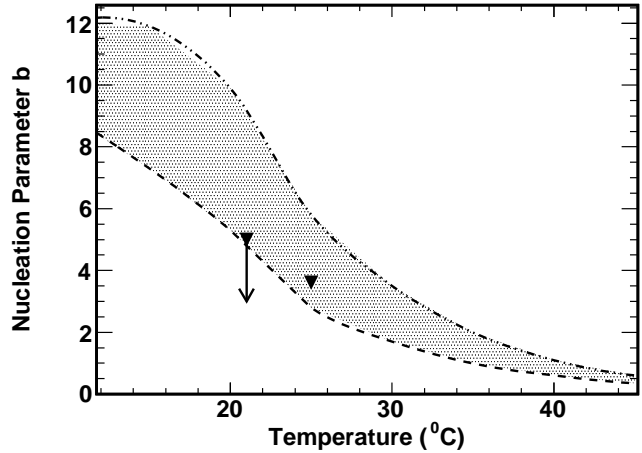}
\end{center}
\caption{Nucleation parameter $b$ as a function of temperature. This parameter uses $L_c$ = $bR_c$ to relate the spatial extension of the energy deposition with the critical radius defined in relation (\ref{eq:1}). If carbon would trigger at threshold, then only nucleation parameters above the upper line would be allowed. If fluorine recoils determine the threshold, then the shaded region between the two curves defines the allowed parameter space. The threshold data of the alpha spiked detectors yield an additional upper limit at 21\degC{} and an additional prediction at 25 \degC{} (triangles); the alpha data favour the lower part of the shaded parameter space.}\label{fig:6}
\end{figure}

In the case of alpha particles, the threshold energy $E_{\mathrm{th}}$(T) for  particles entering from outside the droplets is related to the deposited energy by relation (\ref{eq:3}) where $(dE/dx)_{Bragg}$ = 210 keV/$\mu$m is the maximum specific energy loss at the end of the track of a 5.64 MeV alpha particle emitted in $^{241}$Am decays. From this relation it follows that $L_c$ = 0.31 $\mu$m, for the critical length $L_c$ along which the particle deposits its energy in order to be able to create a critical proto-bubble. Using in addition the prediction by the Seitz model given in (\ref{eq:1}) for the critical radius $R_c$(T), one obtains $L_c$ (24$^{\circ}$) = 3.6 $\times$ $R_c$ (24$^{\circ}$), which yields an estimate of the model parameter $b_\alpha$ (24$^{\circ}$) = 3.6 for this temperature. This value is compared in Figure \ref{fig:6} with those obtained from neutron calibrations.

On the other hand one can use the fact that alpha particles do not trigger phase transitions at 21$^{\circ}$ and E$_{\mathrm{dep}}$ =146 keV  in order to obtain an upper limit on the critical length of  $L_c$(21\degC{})  $\le$  0.45$\mu$m which implies  $b_{\alpha}$ (21\degC{})  $\le$ 5, as indicated in Figure \ref{fig:6}.  

Incidentally, these values of $L_c$ are close to the range of  \flourine{} at this threshold energy, but definitely smaller than the range of \carbon{12} with $R_C$ = 1.16 $\mu$m. Since from Figure 1 the stopping power of carbon is always smaller than that of fluorine, its energy deposition would be around 80 keV and therefore smaller than the 146 keV required, which supports the assumption made in sect. \ref{sec:calibmono} that fluorine triggers at threshold.   

\subsection{L$_c$ from neutron induced recoils}
 Applying relation (\ref{eq:3}) on the data taken during the neutron calibrations reported in sect. \ref{sec:calibmono}, two sets of parameters $b_F$(T) and $b_C$(T) result, depending on the assumption whether either \flourine{} recoils trigger at threshold (lower curve in Figure 6) or \carbon{12} recoils (upper curve in Figure 6).  Both curves represent lower bounds on $L_c$. In particular, in the case of $b_F$ the area between the two curves is allowed, since above that region carbon would trigger at threshold \cite{28}. The two estimates for $L_c$ discussed in the alpha scenario add two independent constraints which favour the lower set of $b_F$ values. In summary, the range of the preferred values, together with the observed trend that $L_c$ increases with temperature agree well with measurements on several other halocarbons discussed in \cite{13}.

\section{Acoustic Signals from Particle Induced Events in Superheated Liquids}\label{sec:acoustic}
 
          It is known that energetic charged particles traversing non-stressed liquids or solids produce acoustic waves during their passage. This so-called thermo-acoustic effect was predicted and described by Askarian et al. \cite{30} and is exploited in high energy neutrino detection in the PeV range by the ANTARES and ICECUBE experiments \cite{31,32}.  However, applied to the processes considered here, with energy depositions in the keV range, the emitted sound intensities predicted by the thermo-acoustic effect are undetectable. On the other hand, particle interactions in stressed or superheated liquids, produce detectable acoustic signals which are related to the nature or the extension of the primary event; this suggests that the phase transition process in superheated liquids provides an intrinsic acoustic amplification mechanism with a gain of at least 10$^{5}$\cite{33}.  

\subsection{Neutron induced recoils}  
\begin{figure}
\begin{center}
 \includegraphics[scale=0.6]{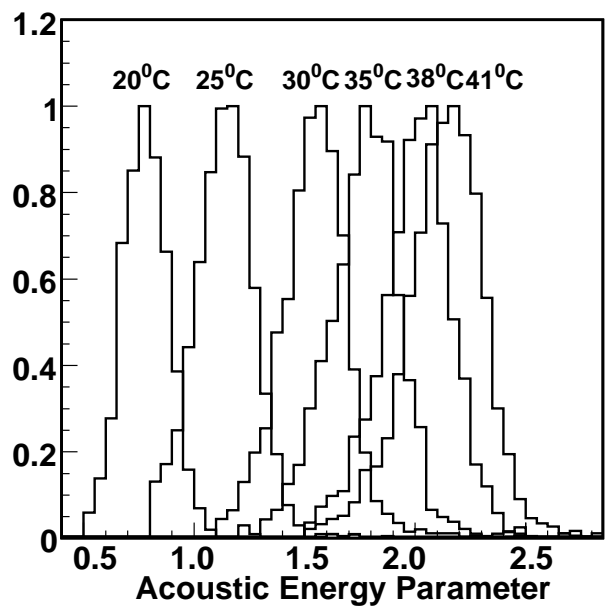}
\end{center}
\caption{Distribution of the acoustic energy parameter recorded in calibrations with poly-energetic neutrons from an AmBe source. For a given event the signal energy is constructed by squaring the amplitudes and averaging over the waveforms of at least six transducers per detector. The logarithm of this quantity is used to define the acoustic energy parameter. Neutron-induced recoils show up in a peak, which is well separated from acoustic and electronic noise.}\label{fig:7}
\end{figure}

Calibration data with fast neutrons of AcBe, AmBe and Cf sources showed that the associated waveforms have a short rise time, reaching the maximum amplitude after 20-40$\mu$s, with slower oscillations following for several milliseconds. In order to characterize signals of different origins in the detector, a Bessel band pass filter is applied to cut off frequencies below 18 kHz and then the waveform of each transducer is squared and integrated over the signal duration, starting from a fixed pre-trigger time.  The resulting values are then averaged over all active transducers for each event in order to mitigate solid angle effects.  The logarithm of this averaged acoustic energy is used to define the acoustic energy parameter, so called as it is a measure of the average energy contained in the transducer signals. It shows a well defined distribution, with a centroid which increases smoothly with temperature \footnote{The values of the acoustic parameters in Figures 7, 8 and 10 differ slightly among each other due to different experimental conditions such as amplifier gains and a steadily refining analysis; still in each graph the same definition is being used for both neutrons and alphas (gammas) and for different temperatures.}(Figure \ref{fig:7}).   

This observation suggests that the fast component of the signal does not depend on the droplet size, but retains information of the very first stage of bubble formation \cite{8}. A plausible explanation of this effect will be given in sect. \ref{sec:bubgrowth}.  This property can be used to discriminate particle induced recoil events from non-particle related signals \cite{5, 34}.  Since WIMP induced recoils are similar to neutron induced recoils, this feature is of prime importance for dark matter searches with superheated liquids.

\subsection{Alpha decays}  Alpha decays in the PICASSO detectors also produce signals with well defined energy, and if fully contained in a droplet, with larger acoustic energy than observed in neutron induced events \cite{6}. This can be explained by the fact that the ranges of neutron induced recoils of keV energies are of sub-$\mu$m extension and therefore comparable in size to the critical length $L_c$. Therefore these events are able to trigger only one primary nucleation. However, alpha emitters located within the superheated liquid can trigger at least two vaporizations: one from the recoiling nucleus and the second one or more on the alpha particle track. In both cases the energy released during vaporisation increases with temperature, but stays well defined for a given temperature. 
Data taken with the alpha spiked detectors described in sect. \ref{sec:calalpha} have been used to investigate this nucleation hypothesis further. In order to detect possible deviations from single bubble nucleations, each alpha measurement at a given temperature was followed by an exposure to an AcBe neutron source.  It was found that for the Am spike, where only alpha particles originating from outside the droplets are able to trigger a phase transition, the signal energy distribution coincided with the distribution recorded during the neutron sessions.
 \begin{figure}
\begin{center}
 \includegraphics[scale=0.8]{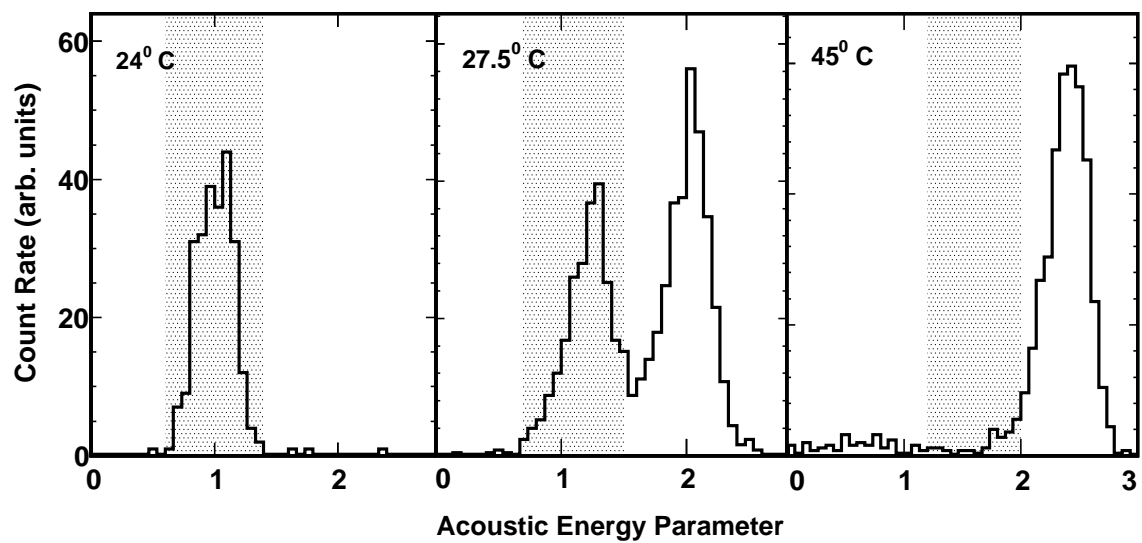}
\end{center}
\caption{Distributions of the acoustic energy parameter observed as a function of temperature with alpha contaminated detectors where the alpha activity occurs inside the droplets. The vertical bands indicate the location of recoil events produced during calibrations with an AcBe-neutron source. Left (24\degC{}): the signal strengths of recoil nuclei in alpha decays coincide with those from neutron calibrations. Centre (27.5 \degC{}): a second peak appears on the high side which is caused by the joint effect of recoil nuclei and the energy deposition by the alpha track. There are still events where only recoils nucleate.  Right (45\degC{}): at this temperature $\alpha$-particles and recoil nuclei contribute simultaneously to the signal.}\label{fig:8}
\end{figure}

\begin{figure}
\begin{center}
 \includegraphics[scale=0.7]{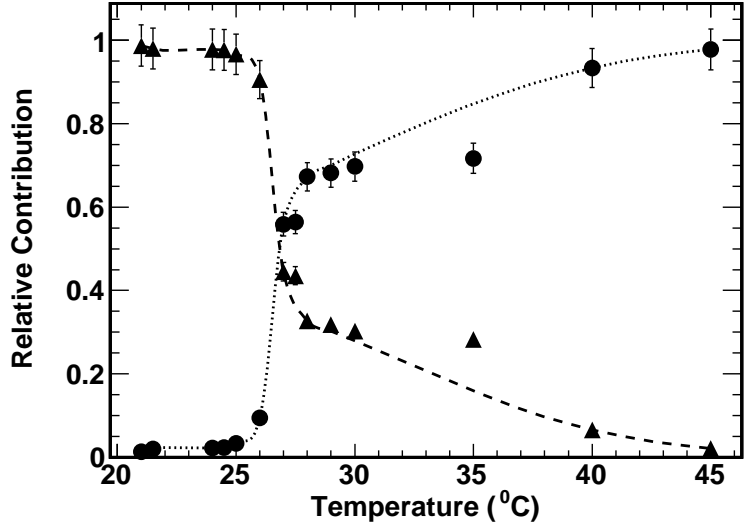}
\end{center}
\caption{The two relative contributions to the acoustic signal energy for alpha contaminations within the droplets: nuclear recoil induced events corresponding to the first peak in Figure 8 (triangles); joint contribution of nuclear recoils \emph{and} alpha particle induced events corresponding to the second peak in Figure 8 (dots). The first peak coincides with the acoustic energy parameter of neutron induced recoils and dominates between 21 and 25\degC{}; above that temperature the detector becomes sensitive also to alpha particles which adds to the strength of the signal, but does not change the count rate, since the detector is already fully sensitive. Curves are drawn to guide the eye. }\label{fig:9}
\end{figure}

  For the $^{226}$Ra spiked detector and regular 4.5L PICASSO detectors with high intrinsic alpha background rates, a different pattern arises when the distribution of the acoustic energy parameter is recorded as a function of temperature (Figure \ref{fig:8}). Between threshold at 21\degC{} and below 25\degC{},  only recoil nuclei create a peak which coincides with the location of the neutron induced recoils during exposure with an AcBe source. With further increase in temperature above 25$^{\circ}$, when the detector becomes sensitive to alpha particles a second peak arises at higher acoustic energy and the first peak gradually diminishes. This second peak corresponds to nucleations caused by  recoil nuclei plus nucleations caused by the Bragg peak on the alpha track. During this redistribution between the peaks, the sum of the count rates remains constant; the relative contribution to the total count rates are shown in Figure \ref{fig:9}. The apparent shift of +2\degC{} between the threshold data shown in Figure 2 and the data shown in Figures 8 and 9 is due to an equivalent 0.2 bar difference in operating pressure: the spiked detectors (Figure 2) were operated at 1 bar surface ambient pressure, whereas the data shown in Figures 8 and 9 were taken at the SNOLAB underground site at 1.2 bar ambient pressure.     
  
It was also noticed that the degree of separation between recoils and alpha particles depends on the temperature and the frequency content of the signals: at temperatures around 25\degC{} high-pass filters which accept frequencies above 10 kHz give the best result and the resolution tends to decrease with increasing cut-off frequency. The opposite happens at high temperatures - above 40\degC{} - where the best results are obtained with cut-offs above 100 kHz and discrimination worsens for lower cut-offs. 

\subsection{Delta-electrons from $\gamma$-ray induced events} 
\begin{figure}
\begin{center}
 \includegraphics[scale=0.6]{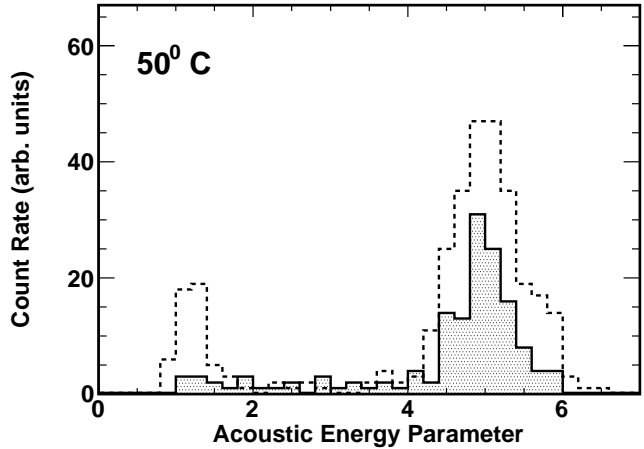}
\end{center}
\caption{Acoustic energy parameter for $\gamma$-induced events. Data were taken in presence of a $^{137}$Cs source (shaded) and compared with the signals from a $\gamma$-shielded AcBe neutron source (dotted line). $\gamma$-induced events are caused by clustered energy depositions (Auger or $\delta$- electrons) on the track of Compton electrons traversing the droplets. The acoustic energy distribution of these events coincides with that of events with single nucleation.
}\label{fig:10}
\end{figure}

If detectors are operated at temperatures far below the plateau-$\gamma$ senistivity (i.e 65\degC{}) then the clustered energy depositions from Auger- or $\delta$- electrons on the tracks of Compton scattered electrons create events with small multiplicity within a droplet (sec. \ref{sec:gamma}). Therefore the acoustic signals are expected to reproduce those produced by single nucleations on the short tracks of nuclear recoils. In order to verify this hypothesis, data were taken with two different detectors in the presence of a $^{22}$Na and a $^{137}$Cs  source, respectively,  and compared with the signals induced by fast neutrons from a $\gamma$-shielded AcBe source. Measurements were performed at 45 and 46 and 50\degC{}, respectively, and the acoustic energies of the $\gamma$-induced signals coincided with those of the neutron induced recoils (Figure \ref{fig:10}). It would be interesting in future to extend these measurements to higher temperatures, where also multiple nucleations might become observable on the Compton electron tracks traversing the droplets (the high temperature data in sec. \ref{sec:gamma} were taken without neutron reference measurements).       

\section{Dynamics of Bubble Growth and Acoustic Signal Formation}\label{sec:bubgrowth}
           The dynamics of bubble growth and the associated sound emission in superheated liquids is a complex phenomenon involving non-linear thermodynamic processes which are still the subject of ongoing research \cite{35}. Particle induced sound generation in superheated liquids was first discussed in \cite{33}, but up to now only an approximative and qualitative description can be given of the processes leading to the recorded acoustic signals and the observed alpha Ð recoil discrimination. 
          The early theoretical works of Rayleigh \cite{36}, Plesset and Zwick \cite{37} were based on an approximate solution of the Navier-Stokes equation and postulated that the growth of a vapour bubble in a superheated liquid is controlled by three stages: 1) a surface tension controlled stage, 2) followed by a stage where the growth is limited by the inertia of the liquid and where  the bubble expands with a constant velocity and 3) an asymptotic stage which is dominated by heat transfer and where the bubble growth is decelerating. 
 
	          As soon as the bubble radius reaches $R > R_c$, the expansion is driven by the energy stored in the bubble itself and its vicinity and this is described by the Rayleigh-Plesset equation \cite{36}. Its solution implies that the radius increases linearly with time and that the speed of this inertial growth is proportional to the square root of the superheat $\Delta p$ defined in sect. \ref{sec:model}:

 \begin{equation}
 	R_{in}(t) = A(T) \times t, \quad\quad\quad\quad\quad\quad A(T) = \left(\frac{2}{3\rho_l}\Delta p\right)^{1/2}
\end{equation}\label{eq:9}

Since the superheat increases and the liquid density decreases with temperature,  the speed of bubble growth also increases with temperature. In particular we find for \cfourften \ a prediction of the growth velocity of $A$(30\degC{}) = 11.6 $\mu$m/$\mu$s and $A$(46\degC{}) = 13.6 $\mu$m/$\mu$s.    
        
       Due to the expansion of the bubble volume the vapour within the bubble and also the liquid in the vicinity of the bubble walls cools down until it reaches the boiling temperature and after a certain characteristic time $\tau$, further growth is only possible if energy is supplied by heat transfer from more and more distant layers of the liquid. From there on, the growth rate becomes limited by thermal diffusion and it decreases continuously. This is described by the Plesset-Zwick equation \cite{37}, which in this regime predicts a much slower increase of the bubble radius proportional to the square root of time:

 \begin{equation}
R_{th}(t) = B(T)\times t^{1/2}, \quad\quad\quad\quad\quad\quad B(T) = \left( \frac{12}{\pi} \kappa \rho_l c_{pl}\right)^{1/2}\frac{T-T_b}{h_{lv}\rho_v}
\end{equation}\label{eq:10}

\noindent Here $\kappa$ is the thermal conductivity of the liquid and $c_{pl}$ is its specific heat; the other quantities are as defined in sect. \ref{sec:model}. The growth parameter $B(T)$ can also be expressed as $B(T) = (3\kappa/2\pi) J_a (T)$, where $J_a$ is the Jakob number, a dimensionless quantity which characterises the speed of the bubble growth.  In particular it was found that the $t^{1/2}$ - law is only valid for $2 <  J_a < 100$ \cite{35,38}. For \cfourften \ and within the temperature range considered here, $J_a$ follows a distribution with a broad peak around 35\degC{} with $J_a$ = 18.5 and which decreases slowly and asymmetrically to $J_a$ = 16.5 at 20\degC{} and $J_a$ = 18 at 50\degC{}, respectively. The growth rate 1 $\mu$s after nucleation is predicted by (10) to be $\approx$ 2.5 $\mu$m/$\mu$s, which is already smaller than the speed of inertial growth. The time $\tau$, which is the time at which the transition between the two asymptotic solutions occurs, is strongly model dependent  and a scope of investigations \cite{35}.  
 
The predictions of growth rates by the Òclassical modelÓ are, however, idealisations and measurements in superheated liquids showed linear growth rates which were substantially slower than the predicted inertial growth, but still larger than thermal growth up to 100 $\mu$s after nucleation \cite{39}.  
 
             After complete phase transition of a droplet, a freely oscillating vapour bubble is formed. The resulting bubble is a harmonic oscillator, oscillating around its equilibrium radius $R_b$ and the ambient equilibrium pressure $P_0$ with a resonance frequency calculated by Minnaert \cite{40} as:

 \begin{equation}
 	\nu_R = \frac{1}{2\pi R_b}\sqrt{\frac{3\kappa P_0}{\rho_l}}
\end{equation}\label{eq:11}

\noindent where $\kappa$ is now  the polytropic coefficient of the gas and $\rho_l$ the density of the surrounding liquid. For \cfourften \ at 30\degC{} the resonance frequency and the bubble radius are related by the simple relation $\nu_R$ (kHz) = 2.4/$R$ (mm). Typical droplets in PICASSO of 100 $\mu$m radius will eventually form bubbles of   $R_b \approx$ 0.35 mm radius and are expected to oscillate with a fundamental frequency of $\nu_R \approx$ 6 kHz. This frequency is below the 18 kHz cut-off used in the analysis of signals discussed in sect. \ref{sec:acoustic}. 
 
The pressure of the emitted sound, which is produced in the liquid  by an expanding or oscillating spherical bubble of radius $R$(t), is related to the acceleration of its volume $V$(t) : 

\begin{equation}
 	\Delta P(r,t) = \frac{\rho_l}{4\pi}\frac{\ddot{V}(t-r/c)}{r} = \frac{\rho_l}{4\pi r}\left(\frac{4}{3}\pi \right) \frac{d^2R^3}{dt^2} = \frac{\rho_l}{r}(2R \dot{R}^2 + R^2 \ddot{R})
\end{equation}\label{eq:12}

\noindent where $\Delta P$($t,r$) is the pressure change produced in the liquid at a distance $r$ from the source, $c$ is the velocity of sound and $\rho_l$ is the density of the liquid \cite{41}.   Inserting the solutions for inertial growth $R_{in}$(t) and for thermal diffusion limited growth $R_{th}$(t)  into (\ref{eq:12}), one finds the radiated pressure signals for the two modes of asymptotic bubble growth: 

\begin{equation}
	\Delta P_{in} \propto \rho_l A(T)^3 \times t \quad\quad\mathrm{and}\quad\quad \Delta P_{th} \propto \rho_l B(T)^3 \times t^{-1/2}
\end{equation}\label{eq:13}

Piezo-electric transducers are sensitive to the instantaneous pressure $\Delta P$, with sensitivities quoted in terms of $\mu$V/$\mu$bar (sect. \ref{sec:det}). Therefore an analysis of the waveform of the transducer signal allows one to obtain information about the emission process. However, given   our present experimental conditions our timing information is severely limited by the sampling frequency (2.5$\mu$s/sample) and is distorted especially for times larger than 80$\mu$s by reflections and container effects, like sound propagation in the acrylic.  
 \begin{figure}
\begin{center}
 \includegraphics[scale=0.7]{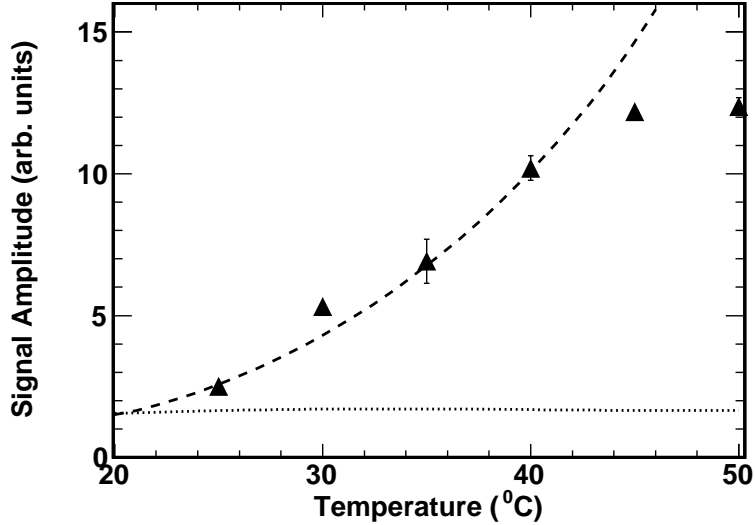}
\end{center}
\caption{Amplitudes of the measured acoustic signals as a function of temperature compared to theoretical predictions. In the case where bubble growth is driven by thermal diffusion only, the observed signal amplitudes decrease slowly with temperature (dotted); inertial growth predicted by the Rayleigh-Plesset solution of the Navier-Stokes equation predicts a steady rise of the pressure signal with temperature and correspondingly also of the amplitudes (broken). The two theoretical predictions were set to a common value at 20\degC{}. }\label{fig:11}
\end{figure}

Nevertheless some conclusions about the bubble growth and acoustic signal production can be inferred from the observed amplitudes and their dependence on temperature.   Figure \ref{fig:11} compares the measured amplitudes as a function of temperature with those predicted by the two growth models. Apparently only the inertial growth scenario in which the amplitudes increase with temperature shows a trend similar to the data.  Since the observed alpha Ð recoil discrimination implies a spatial resolution of two nucleation centres separated by about the length of an  alpha track, i.e. $L_{\alpha} \sim$ 40 $\mu$m,  it can be concluded that the inertial phase cannot last much longer than about 2 $L_{\alpha}$ / $A(T)$ : if inertial growth would continue beyond that time, the expanding bubble volumes would have merged completely and all information about multiple nucleation sites would have been washed out. Therefore after that time, i.e. t $\geq$ 10 $\mu$s, and  according to the above estimates for $A(T)$,  the decelerating thermal growth phase must have become the dominant effect in order to preserve the information about the spatial extension of the original nucleation volume.  
 
           At the moment our alpha - recoil discrimination data indicate the presence of two to three nucleation centres. However, within the scenario discussed above one would expect that with a better timing resolution of the acoustic readout system, more nucleation centres or an extended nucleation region could be resolved along the alpha track at an earlier stage of formation, which would result in a further improvement of the alpha Ðrecoil discrimination.

\section{Conclusions}
              The full sensitivity of superheated liquids to nuclear recoils in the absence of a significant sensitivity to gamma rays or minimum ionising particles has stimulated the interest in this technique for dosimetry, neutron detection in fusion research and recently large scale applications in dark matter searches. However some grey areas exist in the detailed understanding of the underlying radiation detection processes: how precisely does radiation induce phase transitions at the nanometer scale, what are the precise dynamics of bubble growth and what time scales are involved, how are the observed acoustic signals produced, and how much information about the nature of the primary event do they contain? 
 
            Our recent studies were able to consolidate some known features and to shed new light on some of the open questions : 1) the energy thresholds predicted by the classic nucleation theory in \cfourften \ are in good agreement with neutron and alpha calibration data; only at the lowest neutron energy at 4.8 keV does a small discrepancy exist which needs to be clarified; 2) threshold measurements with detectors spiked with alpha emitters allow one to differentiate between energy depositions by the recoiling nuclei and those caused by the Bragg peak of alpha particles; 3) recoil nuclei following alpha decay have a higher energy threshold than alpha particles; 4) signals produced simultaneously by recoil nuclei and alpha particles have more acoustic energy  than signals produced by one or the other separately;  5) neutron and alpha data deliver a consistent picture of how the critical interaction length $L_c$ evolves in terms of the critical radius $R_c$ and with temperature; 6) signal amplitudes increase with temperature which implies that inertial growth contributes to acoustic signal formation; 7) the observed alpha  - recoil discrimination requires a fine tuning in the interplay between the inertial bubble growth mode and the asymptotic thermal growth. While our understanding of this interplay is rudimentary at this stage, it is expected that much better discrimination between particles can be achieved by adapting the speed of the acoustic read-out chain to the timescale which carries most of the information about the primary nucleation process.

\section*{Acknowledgements}
We wish to acknowledge the support of the National Sciences and Engineering Research Council of Canada (NSERC), the Canada Foundation for Innovation (CFI), the National Science Foundation (NSF) for funding and the Czech Ministry of Education, Youth and Sports. We also thank SNOLAB and its staff for its hospitality and for providing help and competent advice whenever needed, as well as Andrea Vollrath, Andrea Palenchar and Brendan Sweeney at IUSB. Special thanks go to prof. Mala Das from the Saha Institute for Nuclear Physics, Kolkata, for helpful and inspiring discussions. This work is also supported by the NSF grant PHY-0856273.

\end{document}